\pdfoutput=1
\documentclass{emulateapj}
\usepackage{natbib}
\usepackage{amsmath}
\usepackage{verbatim}
\usepackage{placeins}

\usepackage{appendix}

\begin{document}

\title{\textbf{Radar imaging and characterization of Binary Near-Earth Asteroid (185851) 2000 DP107}}
\date{}
\author{S. P. Naidu\altaffilmark{1}, J. L. Margot\altaffilmark{1,2},
P. A. Taylor\altaffilmark{3}, M. C. Nolan\altaffilmark{3},
M. W. Busch\altaffilmark{4}, L. A. M. Benner\altaffilmark{5},
M. Brozovic\altaffilmark{5}, J.~D.~Giorgini\altaffilmark{5}, J. S. Jao\altaffilmark{5},
C. Magri\altaffilmark{6}}

\altaffiltext{1}{Department of Earth, Planetary, and Space Sciences,
University of California, Los Angeles, 595 Charles Young Drive East,
Los Angeles, CA 90095, USA} 

\altaffiltext{2}{Department of Physics and
Astronomy, University of California, Los Angeles, 430 Portola Plaza,
Los Angeles, CA 90095, USA} 

\altaffiltext{3}{Arecibo Observatory, HC3 Box 53995, Arecibo, PR 00612, USA}

\altaffiltext{4}{SETI Institute, 189 Bernardo Ave. Suite 100, Mountain View, Mountain View, CA 94043, USA}

\altaffiltext{5}{Jet Propulsion Laboratory, California Institute of Technology, Pasadena, CA 91109-8099, USA}

\altaffiltext{6}{University of Maine at Farmington, 173 High Street, Preble Hall, Farmington, ME 04938, USA}

\begin{abstract}
Potentially hazardous asteroid (185851) 2000 DP107 was the first
binary near-Earth asteroid to be imaged.  Radar observations in 2000
provided images at 75~m resolution that revealed the shape, orbit, and
spin-up formation mechanism of the binary.  The asteroid made a more
favorable flyby of the Earth in 2008, yielding images at 30~m
resolution.  We used these data to obtain shape models for the two
components and to improve the estimates of the mutual orbit, component
masses, and spin periods.  The primary has a sidereal spin period of
$2.7745 \pm 0.0007$ h and is roughly spheroidal with an equivalent
diameter of $863$~m $\pm~5\%$.  It has a mass of
$4.656 \pm 0.43\times10^{11}$~kg and a density of 
$1381 \pm 244$~kg~m$^{-3}$.  It exhibits an equatorial ridge similar
to the (66391) 1999~KW4 primary, however the equatorial ridge in this
case is not as regular and has a $\sim300$~m diameter concavity on one
side.  The secondary has a sidereal spin period of 1.77 $\pm$ 0.02
days commensurate with the orbital period.  The secondary is slightly
elongated and has overall dimensions of $377\times314\times268$~m (6\%
uncertainties).  Its mass is $0.178 \pm 0.021\times10^{11}$~kg and its
density is $1047 \pm 230$~kg~m$^{-3}$.  The mutual orbit has a
semi-major axis of $2.659 \pm 0.08$~km, an eccentricity of $0.019 \pm
0.01$, and a period of $1.7556 \pm 0.0015$ days.  The normalized total
angular momentum of this system exceeds the amount required for the
expected spin-up formation mechanism.  An increase of angular momentum
from non-gravitational forces after binary formation is a possible
explanation.
The two components have similar radar reflectivity, suggesting a
similar composition consistent with formation by spin-up.  The
secondary appears to exhibit a larger circular polarization ratio than
the primary, suggesting a rougher surface or subsurface at radar
wavelength scales.
\end{abstract}

\section{Introduction}
Asteroid (185851) 2000 DP107 was discovered on 2000 February 29 by the
Lincoln Near-Earth Asteroid Research (LINEAR) program in New
Mexico. Radar observations in October of that year revealed the asteroid
to be a binary system~\citep{margot02}, the first such system to be
imaged in the near-Earth asteroid (NEA) population.  The radar data
were instrumental in establishing that NEA satellites form by a
spin-up and rotational fission process~\citep{margot02}. Additional
radar and photometric studies showed that $\sim15\%$ of all NEAs
larger than 200~m are binary in
nature~\citep{pravec99,margot02,pravec06}.  For a recent review of the
properties of binary asteroids, see~\citet{marg15AIV}.

The presence of a satellite around the primary gives us an opportunity
to secure direct measurements of several quantities that are not
normally measurable.  Radar observations enable calculations of the
orbital period and orbital separation, which reveal the total mass of
the binary system through Kepler's third law.  In addition, radar
observations enable measurements of the masses of individual
components by measuring the distances of the component centers of mass
(COMs) from the system COM.  Using this information along with the
shape models of the two components derived from radar images, we can
estimate their densities.  These are important constraints for testing
models of formation and evolution of NEAs.

Radar observations of 2000 DP107 in 2000 October yielded rough
estimates of masses, sizes, and densities of the two components as
well as their mutual orbit~\citep{margot02}.  In 2008 September, the
asteroid made another close approach to the Earth and was observed at
a distance $\sim0.06$ astronomical units (AU) or about 20 lunar
distances.  Because this was about half the distance of the 2000
encounter ($\sim0.11$~AU), it resulted in data sets with
signal-to-noise ratio (SNR) $\sim20$ times higher than in 2000.  The
high SNRs enabled us to derive component shapes with effective
resolutions of $\sim50$~m on the surface, and estimate the component
masses, volumes, and densities more accurately than was possible in
\citet{margot02}.

In this paper we present detailed component shape models, improved
estimates of component masses and densities, and estimates of the
mutual orbit parameters using the 2000 and the 2008 radar data.  This
detailed characterization of 2000 DP107 and its favorable
accessibility ($\Delta v \approx 5.9$ km s$^{-1}$) make it a good
candidate for spacecraft rendezvous missions.
2000 DP107 was the target of the PROCYON mission~\citep{funase14},
which had a planned flyby the asteroid in 2016. However the mission
suffered an engine failure and will be unable to perform the flyby.

\section{Methods}

\subsection{Observing and Data Processing}
\label{sec:observing}
\renewcommand{\thefootnote}{\fnsymbol{footnote}} We observed 2000
DP107 using the Arecibo S-band (2380 MHz, 13 cm) radar and the
Goldstone X-band (8560 MHz, 3.5 cm) radar on 10 days between 2008
September 9 and 24, during which the asteroid moved $\sim60^\circ$
across the sky. It came closest to Earth on September 11 at a distance
of 0.057~AU.  Most of the observing time was dedicated to
range-Doppler imaging, with the remainder dedicated to collecting
continuous wave (CW) spectra\footnote[1]{The word continuous is used
to distinguish this transmission mode from range modulated operation.}.  We
obtained 335 range-Doppler images and 65 CW spectra using Arecibo and
534 range-Doppler images and 67 CW spectra using Goldstone.

Radar observations were carried out according to the methods described
in \citet{naid13}.  Briefly, radar imaging was carried out by
transmitting a repeating pseudo-random code modulated over a
circularly polarized carrier wave, using a binary phase shift keying
scheme~\citep{proakis2007}. In each {\em run}, the waveform was
transmitted for approximately the round-trip light-time (RTT) before
switching over to the receiver. The received signal was demodulated
and then decoded by cross-correlating it with a replica of the
transmitted code, yielding a range resolution equal to the baud length
of the transmitted code.  In each range bin, consecutive returns were
fast Fourier transformed (FFT) to obtain the received signal power as
a function of Doppler frequency.  The end product is a two-dimensional
array or image showing the echo power as a function of range and
Doppler frequency.
Technically, the observable measured at the telescope is the
round-trip light time to the target.  We obtained range bins by
multiplying the baud length of the code by half the speed of light.

For CW runs, a monochromatic wave was transmitted for the RTT to the
asteroid before switching over to the receiver. The received signal
was demodulated, sampled, and recorded. A FFT was applied to the echo
timeseries to obtain the CW spectra.

Table~\ref{tab:observingsummary} summarizes our observations.  Because
of the smaller antenna size and transmitter power, the Goldstone data
have much lower SNRs ($\sim 1/20$) compared to the Arecibo data.  Six
to eight consecutive Goldstone runs were summed incoherently in order
to improve the SNR.

\begin{deluxetable*}{ccccccccccccc}
\tablewidth{0pt} 
\tablecaption{Radar Observations of (185851) 2000 DP107 in 2008}
\tablehead{ 
\colhead{Tel} & \colhead{UT Date} & \colhead{Eph} & \colhead{RTT} & \colhead{P$_{\rm tx}$} & \colhead{Baud} & \colhead{Prim. res.} & \colhead{Sec. res.} & \colhead{Code} & \colhead{Start-Stop} & \colhead{Runs}\\
\colhead{} & \colhead{yyyy-mm-dd} & \colhead{} & \colhead{s} & \colhead{kW} & \colhead{$\mu$s} & \colhead{Hz} & \colhead{Hz} & \colhead{} & \colhead{hhmmss-hhmmss} & \colhead{}
} 
\startdata

G     & 2008-09-09 &  85  &  59 & 445 & 1.0 &     &     & 127   & 114323-114719 & 3\\
      & $ $        &  87  &     & $ $ & 1.0 &     &     & 127   & 121534-152227 & 96\\
  $ $ & $ $        &  $ $ & $ $ & $ $ & cw  & 1.0 &     & none  & 153128-154316 & 7\\
\\		   		      		  								  
\hline\\	   		      		  							  
G     & 2008-09-10 &  89  &  58 & 445 & cw  & 1.0 &     & none  & 100115-100907 & 5\\
      & $ $        &  $ $ &     & $ $ & 0.5 &     &     & 8191  & 102915-152812 & 153\\
  $ $ & $ $        &  $ $ & $ $ & $ $ & cw  & 1.0 &     & none  & 153412-154908 & 7\\
\\		   		      		  								  
\hline\\	   		      		  							  
A     & 2008-09-10 &  89  &  58 & 628 & cw  & 0.2 &     & none  & 101757-102635 & 5\\
  $ $ & $ $        &  $ $ & $ $ & 611 & 0.2 & 0.08& 0.04& 65535 & 102829-114008 & 37\\
  $ $ & $ $        &  $ $ & $ $ & 561 & cw  & 0.2 &     & none  & 114159-115234 & 6\\
  \\
\hline\\
A     & 2008-09-11 &  89  &  58 & 630 & cw  & 0.2 &     & none  & 094357-095235 & 5\\
  $ $ & $ $        &  $ $ & $ $ & 616 & 0.2 & 0.08& 0.04& 65535 & 095452-112810 & 48\\
  $ $ & $ $        &  $ $ & $ $ & 580 & cw  & 0.2 &     & none  & 113030-114115 & 6\\
  \\
\hline\\
G     & 2008-09-12 &  89  &  58 & 430 & cw  & 1.0&      & none  & 095102-100843 & 10\\
      & $ $        &  $ $ & $ $ & $ $ & 0.5 &     &     &  8191  & 101952-142936 & 128\\
  $ $ & $ $        &  $ $ & $ $ & $ $ & cw  & 1.0 &     &  none  & 143725-145705 & 11\\
\\								  								   
\hline\\							  							   
								  								   
A     & 2008-09-13 &  89  &  59 & 603 & cw  & 0.2 &     &  none  & 084405-085301 & 5\\
  $ $ & $ $        &  $ $ & $ $ &     & 0.2 & 0.08& 0.04&  65535 & 085548-105703 & 57\\
  $ $ & $ $        &  $ $ & $ $ & 570 & cw  & 0.2 &     &  none  & 105927-110823 & 5\\
  \\					      		  								   
\hline\\				      		  							   
G     & 2008-09-13 &  89  &  59 & 432 & cw  & 1.0 &     &  none  & 091304-092304 & 5\\
  $ $ & $ $        &  $ $ & $ $ & $ $ & 1.0 &     &     &  8191  & 103240-124441 & 67\\
  $ $ & $ $        &  $ $ & $ $ & $ $ & cw  & 1.0 &     &  none  & 125127-131326 & 12\\
\\					      		  								   
\hline\\				      		  							   
G     & 2008-09-14 &  89  &  60 & 432 & cw  & 1.0 &     &  none  & 092211-094047 & 10\\
  $ $ & $ $        &  $ $ & $ $ & $ $ & 1.0 &     &     &  8191  & 095318-112614 & 47\\
  $ $ & $ $        &  $ $ & $ $ & $ $ & 1.0 &     &     &  8191  & 114209-130244 & 40\\
\\
\hline\\
A     & 2008-09-15 &  89  &  63 & $\sim604$ & cw  & 0.2 &     & none  & 075347-080310 & 5\\
  $ $ & $ $        &  $ $ & $ $ & $ $ & 0.2 & 0.08& 0.04& 65535 & 080616-102000 & 56\\
  $ $ & $ $        &  $ $ & $ $ & 585 & cw  & 0.2 &     & none  & 102236-102952 & 4\\
  \\  
\hline\\
A     & 2008-09-18 &  89  &  70 & 595 & cw  & 0.2 &     & none  & 065646-070944 & 6\\
  $ $ & $ $        &  $ $ & $ $ &     & 0.5 & 0.24&     & 8191  & 071531-092200 & 54\\
  $ $ & $ $        &  $ $ & $ $ &     & cw  & 0.2 &     & none  & 092751-093340 & 3\\
  \\								  								  
\hline\\							  							  
A     & 2008-09-21 &  89  &  81 & 590 & cw  & 0.2 &     & none  & 060205-061410 & 5\\
  $ $ & $ $        &  $ $ & $ $ & 624 & 0.5 & 0.24&     & 8191  & 062313-082654 & 45\\
  \\					      		  								  
\hline\\				      		  							  
A     & 2008-09-24 &  89  &  93 & 660 & cw  & 0.2 &     & none  & 052205-053607 & 5\\
  $ $ & $ $        &  $ $ & $ $ & 680 & 1.0 &     &     & 8191  & 053910-073632 & 38\\
  $ $ & $ $        &  $ $ & $ $ & 605 & cw  & 0.2 &     & none  & 073910-075312 & 5\\
\\
\enddata 

\tablecomments{The first column indicates the telescope: Arecibo (A)
  or Goldstone (G). Eph is the ephemeris solution number used.  RTT is
  the round-trip light-time to the target.  P$_{\rm tx}$ is the
  transmitter power. Baud is the delay (i.e., range) resolution (Bauds
  of 0.2, 0.5, and 1 $\mu$s correspond to range resolutions of 30, 75,
  and 150~m respectively). Prim. res. and Sec. res. are the frequency
  (i.e., Doppler) resolutions of the processed data for the primary
  and secondary shape modeling respectively.  Note that the Doppler
  spread of the target scales linearly with the transmitter frequency.
  Code is the length of the pseudo-random code used.  The time-span of
  the received data are listed by their UT start and stop times.  The
  last column indicates the number of runs acquired in each
  configuration.  }
\label{tab:observingsummary}
\end{deluxetable*}

\subsection{Mutual Orbit}
\label{sec:mutualorbit}
We used a least squares procedure similar to that used in both
\citet{margot02} and \citet{ostro06} to fit Keplerian orbits to the
positions of the secondary COM with respect to the primary COM.  We
used data from 2000 October and 2008 September, and initialized the
fitting procedure with thousands of distinct initial conditions
spanning the entire range of plausible values for all orbital
parameters.
From the 2000 data, we selected
2-4 images on each day
from September 30 to October 7, or 20 measurement epochs spanning 8
days, yielding 20 range separations and 20 Doppler separations.  For
the 2008 data set, we measured the component COM separations in 10
Arecibo images on each day of Arecibo observations, and in 6, 3, 6, 2,
and 8 Goldstone images on September 9, 10, 11, 12, and 13
respectively, giving us a total of 95 measurement epochs spanning 16
days, or 190 measurements
(95 range separations and 95 Doppler separations).

We obtained range-Doppler separations between the primary and
secondary COMs using two different techniques. In the first approach
we estimated the COM locations in the images by measuring the
positions of the leading and trailing edges of the components.  In the
second approach we relied on shape models obtained with the {\tt shape}
software~\citep{hudson93,magri07} to locate the component COMs.  If
the shape models are accurate, the second technique can yield superior
estimates of the COM positions, and therefore of the mutual orbit
parameters.  Sections~\ref{sec:primshape} and~\ref{sec:secshape}
describe shape modeling details.

For the first, edge-based approach, we defined leading edges (LE) and
trailing edges (TE) in the images on the basis of a 3$\sigma$ signal
threshold, where $\sigma$ is the standard deviation of the background
noise.  The LE was defined as the first range bin where the object had
a signal higher than 3$\sigma$ whereas the TE was defined as the last
range bin where at least half of the pixels along the Doppler extent
exceeded the 3$\sigma$ threshold. 
The location of this threshold depends on Doppler resolution (shown in
Table~\ref{tab:observingsummary}).
The primary and secondary were assumed to be roughly spherical and
their radii were estimated from radar images to be roughly 450~m and
150~m respectively. With these assumptions the range coordinates of
the component COMs were taken to be 450~m and 150~m behind their
respective leading edges. The Doppler coordinates of the COMs were
assumed to be located in the middle of the Doppler extent on the
trailing edge.  Conservative uncertainties of 2-3 times the range and
Doppler resolutions were assigned to the range-Doppler separations.

For the second, {\tt shape}-based approach, we used the shape modeling
software to locate the component COMs under a uniform density
assumption.  {\tt shape} aligns the synthetic radar images derived from the
shape models with the observed radar images, and outputs the COM
positions used for the alignment with sub-pixel precision.
Uncertainties on the order of the image resolution were assigned to
the {\tt shape}-based range and Doppler separations.  We computed the COM
separations at the same epochs as those used in the edge-based
approach.

We fit the mutual orbit and component shapes in an iterative manner.
Each iteration started with mutual orbit fitting followed by component
shape modeling.  In the first iteration, we used the edge-based
approach to determine the preliminary mutual orbit and used the orbit
solution 
(orbit pole and longitude of pericenter) 
to inform our component shape modeling (Sections \ref{sec:primshape}
and \ref{sec:secshape}). For the second iteration, the best-fit
component shapes from the first iteration were used to refine the COM
separation estimates using the {\tt shape}-based approach. These improved
primary-secondary separation estimates were used to refine the mutual
orbit fit.
The refined mutual orbit solution was used to obtain the final shape
models of the components.

\subsection{Primary Shape}
\label{sec:primshape}
We used the {\tt shape} software~\citep{hudson93,magri07} to invert
the sequence of range-Doppler images and CW spectra from 2008 to
obtain a 3D shape model for the primary.  Our data set consisted of
278 Arecibo range-Doppler images and 95 CW spectra from both Arecibo
and Goldstone covering a 16-day period between 2008 September 9 and
24.  We left out the low-resolution Arecibo images from September 24
and all the Goldstone images as they did not improve the quality of
the fit and slowed down the shape modeling process.  Because we
modeled the primary and secondary separately, we edited the images and
spectra to exclude the contribution of the other component to the
echoes.

Shape modeling was generally carried out in three steps.  First we fit
a triaxial ellipsoid model to the data to get the overall extents of
the object. We then moved on to a 8th-degree-and-order spherical
harmonics model to fit for the 
global-scale topography
seen in the
images.  Finally, in order to fit for the small-scale features, we
used a vertex model with 1000 vertices and 1996 triangular facets.  
This choice yields a facet resolution of $\sim50$~m, which is
comparable to the best range resolution.
In each step weighted penalty functions were used to favor models
having uniform density, principal axis rotation, and a reasonably
smooth surface.  We used a cosine law to model the radar scattering
from the surface of the asteroid:
\begin{equation}
\frac{d\sigma}{dA}=R(C+1)(\cos{\alpha})^{2C}. 
\end{equation}
Here $\sigma$ is the radar cross section, $A$ is the target surface
area, $R$ is the Fresnel reflectivity, $C$ is a parameter related to
the near-surface roughness of the asteroid at the radar wavelength
scales, and $\alpha$ is the incidence angle of the wave.  Values of
$C$ close to 1 represent diffuse scattering, whereas larger values
represent more specular scattering~\citep{mitchell96}.

Because non-linear least-squares methods tend to find local minima
when searching a wide parameter space, we carried out an extensive
grid search for the best-fit spin axis orientation during the
ellipsoid and spherical harmonics shape modeling stages.
We assumed that the lightcurve period of 2.775
h~\citep{pravec06} provided a good approximation to the sidereal
spin period and we fit shape models to the data using spin axis
orientations in increments of 15$^\circ$ in ecliptic longitude
($\lambda$) and 15$^\circ$ in ecliptic latitude ($\beta$).  
For each case, we performed an ellipsoid model fit followed by a
8-degree spherical harmonics model fit. Only the shape parameters,
semi-axes in the ellipsoid fit and spherical harmonic coefficients in
the spherical harmonic fit, the initial rotational phase of the
object, and the radar scattering parameter $R$ were allowed to change.
The spin rate, the spin axis orientation, and the radar scattering
parameter $C$ were kept fixed.  The grid search was repeated for
$C$=0.6, 0.8, 1.0, and 1.2.
We defined a somewhat arbitrary threshold separating acceptable fits
from poorer solutions by visually comparing the synthetic and observed
images and using a $\chi_\nu^2$ threshold of 0.665.

Our mutual orbit pole estimates lie in the region where spin axis
orientations were considered acceptable on the basis of the shape
model fits.  Because we did not obtain a tight constraint on our spin
pole using the shape model search (Section~\ref{sec:rprimshape}), we
used the best-fit mutual orbit pole as the preferred spin pole for
shape modeling.  For a binary formed by a spin-up process one would
expect the primary spin pole to be roughly aligned with the mutual
orbit pole, and tidal processes are expected to damp any residual
inclination.  With this spin pole assumption we fit
8th-degree-and-order spherical harmonics models to the data in the
same way as we did in the grid search. Here we tried values of
sidereal spin rate ranging from 3111~$^\circ$/day ($P=2.777$~h) to
3117~$^\circ$/day ($P=2.772$~h) in steps of 0.2~$^\circ$/day and
values of $C$ ranging from 0.5 to 1.5 in steps of 0.1.  As explained
in Section~\ref{sec:mutualorbit}, the shape modeling was done
iteratively 
with the mutual orbit fits: Mutual orbit fits were
followed by shape model fits.  In the second/final iteration, we
performed the spherical harmonics shape model fit followed by a vertex
model fit.  At each step, we verified the quality of the fit by
visually comparing the synthetic data generated by {\tt shape} with the
corresponding observed data.  For the vertex model fit we used as
initial conditions the best-fit spin state and spherical harmonics
shape model determined at the previous step.  Once again, only the
shape parameters (location of the vertices), the initial rotational
phase, and the radar scattering parameter $R$ were allowed to change
and all the other parameters were kept fixed.

\subsection{Secondary Shape and Spin State}
\label{sec:secshape}
Shape modeling of the
secondary component was performed using a method similar to the one
described in Section~\ref{sec:primshape}.  The data set for modeling
the shape and rotation of the secondary consisted of 180 Arecibo
images taken between 2008 September 10 and 15.  We left out images
with 75~m range resolution from September 18 and 21 because the
secondary was barely resolved in these images, and because these
images did not improve the quality of the shape model fits.  This time
it was the primary that was edited out of the images. The CW
spectra were not used because 
we were not able 
to completely remove the contribution of the primary
from the total echo power.  We fit an ovoid shape model\footnote[2]{An
ovoid is a distorted triaxial ellipsoid such that it has a wide and a
narrow end.}, followed by a 5th-degree-and-order spherical harmonic
model.  We then fit a vertex model with 150 vertices and 296 facets.
\renewcommand{\thefootnote}{\arabic{footnote}}

Periodicities detected in photometric data suggest that the secondary
spin period may be close to 1.76 days~\citep{pravec06}.  This can be
used as a guide in our shape modeling process, being mindful that
lightcurve periods are neither sidereal nor synodic, whereas the {\tt shape}
software 
uses
sidereal spin periods.  This periodicity is close
to the 1.7556 day orbital period (Table~\ref{tab:mutualorbit}),
confirming the finding that the secondary is locked in a 1:1
spin-orbit resonance~\citep{margot02}.  We used the radar-derived,
sidereal orbital period as the nominal spin period of the secondary
for the purpose of shape modeling.  

As with the primary, a grid search did not lead to a conclusive result
about the spin axis orientation.  
Absent recent perturbations, one expects the spin pole of a tidally
evolved secondary to be closely aligned with the mutual orbit pole,
and we used the mutual orbit pole as the spin pole of the secondary.
This proximity to the orbit pole can be verified by computing the
obliquity of Cassini state 1~\citep{peale69}, which is the state
towards which tides drive the satellite spin pole. The other Cassini
states are either unstable or the spin of satellite is unstable at
those Cassini states \citep{gladman96}.  The obliquity can be computed
by using the following equation derived from \citet{gladman96} for a
synchronous secondary:
\begin{equation}
\frac{3}{2}\Big(\frac{C-\frac{A+B}{2}}{C}\Big)
\Big[\frac{\sin{\theta}\cos{\theta}}{\sin{(\theta \pm i)}}\Big]
=\Big(\frac{\dot\Omega}{\omega}\Big).
\label{eq:cassini_prec}
\end{equation} 
Here $A<B<C$ are the the principal moments of inertia of the
secondary, $\theta$ is the obliquity of the secondary spin pole with
respect to the mutual orbit pole, $i$ is the inclination of the mutual
orbit with respect to the invariable plane (in this case it is
approximately the equatorial plane of the primary), $\dot\Omega$ is
the precession rate of the mutual orbit, and $\omega$ is the spin rate
of the secondary.  
After the mutual orbit and the primary and the secondary shapes were
fit (Sections~\ref{sec:rmutualorb},~\ref{sec:rprimshape}, and
\ref{sec:rsecshape}), we 
evaluated Equation~\ref{eq:cassini_prec} with the relevant values and
found that $\theta<1^\circ$, confirming that the expected obliquity is
small.

The initial rotation phase was set to a value such that the secondary
was oriented with its minimum moment of inertia (MOI) principal axis
pointing towards the primary at pericenter.  This is the expected
configuration of a tidally locked satellite.
The same radar scattering law as the one used for the primary was
used.  We allowed the shape parameters and the radar scattering
parameter $R$ to change. The spin rate, the spin axis orientation,
and the radar scattering parameter $C$ were kept fixed.  During the
ovoid model stage, we attempted shape model fits with values of spin
period ranging from 1.6 to 1.9 days in steps of 0.01 days.

An elongated and synchronous secondary in an eccentric orbit about the
primary exhibits librations, which are oscillations about uniform
rotation~\citep[eg.,][]{murray99}. A tidally evolved satellite is
expected to exhibit a relaxed-mode libration~\citep{naidu15}, which
is equivalent to a forced libration~\citep{murray99} when the
spin-orbit coupling is negligible. This libration is roughly
sinusoidal for small eccentricities and its amplitude as a function of
the satellite elongation was estimated by \citet{naidu15} using
numerical simulations.

Because the mutual orbit is eccentric 
($e \simeq 0.02$)
and the secondary is elongated
(Sections~\ref{sec:rmutualorb} and \ref{sec:rsecshape}), we allowed
for the possibility of relaxed-mode libration in longitude in the
rotational model.  For small amplitudes of forced librations and small
orbital eccentricities, the deviation of the secondary orientation
from regular circular motion ($\delta\phi$) can be approximated as
\begin{equation}
\delta\phi\approx A_{lib}\sin{[\omega_{\rm f}(t-t_0)+\pi]},
\label{eq:rotmod}
\end{equation}
where $A_{lib}$ is the amplitude of the forced librations, $\omega_{\rm f}$ is
the forcing frequency which is equal to the mean orbital motion
($n=2\pi/P$), and $t_0$ is the time of pericenter passage.  The
additional phase of $\pi$ appears because for a synchronous secondary
whose natural libration frequency is smaller than the forcing
frequency, the librational phase is expected to be $180^\circ$ at
pericenter~\citep{murray99}.  
We repeated
the ovoid and spherical harmonics shape model fits with libration
amplitudes ranging from 0$^\circ$ to 10$^\circ$ in steps of
$1^\circ$. We tried all possible libration phases in steps of
$4^\circ$ in order to cover the libration phase uncertainty which
arises due to an almost circular orbit. The libration amplitudes and
phases were held at fixed values in each of these fits.

Results of the secondary shape and spin state modeling are discussed
in Section~\ref{sec:rsecshape}.

\subsection{Radar Scattering Properties}
We transmitted circularly polarized waves and used two separate
channels to receive echoes 
with the same circular (SC) and opposite
circular (OC) polarization as that of the transmitted wave
\citep{ostro93}. 
We summed consecutive Arecibo CW runs from 2008
(Table~\ref{tab:observingsummary}) and 
measured the power received in the OC and SC channels.  
The ratio of the power received in SC to the power received in OC
yields the circular polarization ratio which is often denoted by
$\mu_C$.  We also used Equation 1 of \citet{ostro93} to compute the
radar cross-section of the target, 
which has dimension of surface area.
We computed the 
dimensionless {\em specific radar cross-section} ($\Hat{\sigma}$),
also called the {\em radar albedo}, 
with the OC CW spectra by taking
the ratio of the radar cross-section to the geometric cross-sectional
area of the target (primary + secondary)
at the time of observations. We used {\tt shape} to compute the 
orientations of the target and corresponding projected areas at the
times of CW runs.  

The procedure described in the previous paragraph yielded values of
$\Hat{\sigma}$ and $\mu_C$ that combine the echoes from both the
primary and secondary.  We were also interested in estimates of these
quantities for the secondary component alone.  We obtained these by
removing the contribution of the primary from the OC CW spectra.  This
subtraction was performed by fitting a 5th degree polynomial to the
primary CW spectra and by masking out the frequency bins that
contained contribution from the secondary.  After subtraction, we
estimated $\Hat{\sigma}$ and $\mu_c$ for the secondary using the same
procedure as that described in the previous paragraph.  Results are
given in Section~\ref{sec:rradarscat}.

\subsection{Mass Ratio, Component Masses, and Densities}
\label{sec:massratio}
The COMs of the two components follow roughly Keplerian orbits around
the system COM, while the system COM or barycenter orbits the Sun.
The motion of the primary COM relative to the system COM is called the
reflex motion of the primary.  We estimated the mass ratio of the
components and the reflex motion of the primary by quantifying the
goodness of fit of heliocentric orbit fits using astrometry of the
system COM under various mass ratio assumptions.

The system COM lies on the line joining the component COMs at a
distance of $d_p$ from the primary and a distance of $d_s$ from the
secondary.  The ratio of these distances ($d_s/d_p$) is equal to the
primary-to-secondary mass ratio ($M_p/M_s$).  For a given mass ratio
assumption, we calculated the ratio $d_s/d_p$ and estimated the
barycenter locations along the lines joining the component COMs in
each of the 278 images obtained in 2008 that were used for shape
modeling.  This provided estimates of the two-way ranges to the system
COM, where we once again used the {\tt shape}-based component COMs
determined to sub-pixel accuracy.  We explored mass ratio assumptions
from $M_p/M_s$=15 to 30 in steps of 0.1 to determine the corresponding
two-way ranges to the system COM and assigned uncertainties equal to
the range resolution.  For each mass ratio assumption we then
performed a fit for the heliocentric orbit to all available optical
astrometry and the system COM ranges.  The best overall fit, as
indicated by the lowest sum of squares of residuals, yielded an
estimate of the actual mass ratio of the system.

We used the mass ratio to apportion the total mass of the system,
estimated from the mutual orbit, to the primary and the
secondary. These mass estimates were divided by the corresponding
component volume estimates, obtained from shape models, to yield
component density estimates.

\subsection{Primary Gravitational Environment}

We used the primary shape model and density estimate to compute the
gravity field on the surface of the primary,
under a uniform density assumption.
The acceleration on the
surface is the vector sum of the gravitational acceleration due to the
primary's mass and the centrifugal acceleration due to its spin. An
acceleration vector was computed at the center of each facet using the
method described in \citet{werner97}.
The gravitational slope, which is the angle that the acceleration
vector makes with the local inward-pointing surface-normal vector, was
also computed for each facet.

\section{Results}
\label{sec:randd}

\subsection{Mutual Orbit}
\label{sec:rmutualorb}

Our shape modeling results showed that the oblateness $J_2$ of the
primary is about $0.03$ (Section \ref{sec:rprimshape}), such that the
difference between observed and osculating orbital
elements~\citep{gree81} is small.  Specifically, the quantity
$\frac{3}{2} J_2 (R_{\rm p}/a)^2$ ($R_{\rm p}$ is the primary radius
and $a$ is the semimajor axis), which represents the fractional
difference between the observed and osculating values of the
semi-major axis~\citep{gree81}, amounts to $\sim 10^{-3}$.  If the
orbital eccentricity exceeds this value, one can expect an orbital
regime where the true and mean anomalies circulate while the longitude
of pericenter precesses.  For smaller values of the eccentricity,
another class of orbit is possible, where the true and mean anomalies
librate around pericenter while the longitude of pericenter
circulates.  For our purposes, both orbit types are well accommodated
by fitting the observations to a Keplerian ellipse.  However, the
orientation of the ellipse may be different for the 2000 and 2008
observations.  For reasons explained in Section~\ref{sec:dmutualorb},
it was not possible to reliably fit for the apsidal precession rate.

The mutual orbit has a semi-major axis $a=2.659\pm0.08$~km and a
sidereal orbital period $P=1.7556\pm0.0015$~days.  Kepler's third law
yields $GM_T=32.24\pm3.00$~m$^3$~s$^{-2}$, where $G$ is the
gravitational constant and $M_T$ is the total mass of the system.
Substituting $G=6.67\times10^{-11}$~m$^3$~kg$^{-1}$~s$^{-2}$, we find
$M_T=4.834\pm0.45\times10^{11}$~kg.  Table~\ref{tab:mutualorbit} lists
the best-fit orbital parameters obtained using the combined 2000 and
2008 data and compares it to the values published in \citet{margot02}.
The values from both works are consistent with each other.

\begin{deluxetable}{lrr}
\tablecaption{Mutual orbit parameters for 2000 DP107}
\tablehead{
\colhead{Parameter} & \colhead{Value from} & \colhead{Value from}\\
\colhead{} & \colhead{\citet{margot02}} & \colhead{this work}}
\startdata
Semi-major axis (km)                      &  $2.62 \pm 0.16$     &  $2.659 \pm 0.08$ \\
Period (days)                             &  $1.755 \pm 0.007$   &  $1.7556 \pm 0.0015$ \\
Eccentricity                              &  $0.01\pm 0.01$      &  $0.019 \pm 0.01$ \\
System mass ($\times10^{11}$kg)           &  $4.6 \pm 0.5$       &  $4.834 \pm 0.45$ \\
Orbit pole ($\lambda$, $\beta$) ($^\circ$) &  (283, 67) $\pm$ 10 & (294, 78) $\pm$ 10\\
Reduced $\chi^2$                      &  0.32                &  0.23       
\enddata
\label{tab:mutualorbit}
\end{deluxetable}

\subsection{Primary Shape and Spin State}
\label{sec:rprimshape}

\begin{figure}
\plotone{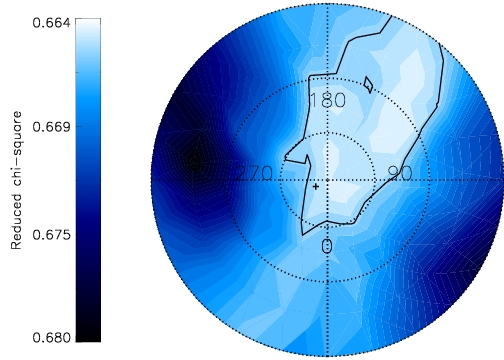}
\caption{Contour plot of goodness of fit ($\chi_\nu^2$) of shape
  models with different spin axis orientations on a polar
  stereographic projection of the celestial sphere, looking down the
  ecliptic north pole. Numbers indicate ecliptic longitudes ($\lambda$).
  Dotted circles (outside to inside) show latitudes 0$^\circ$,
  30$^\circ$, and 60$^\circ$.
   Region enclosed by solid black contour line
  ($\chi_\nu^2=0.665$) shows acceptable shape model fits. Plus sign
  shows our mutual orbit pole estimate.}
\label{fig:shapechi2}
\end{figure}

The result of our grid search for the best-fit spin pole is
illustrated in Figure~\ref{fig:shapechi2}, which shows a contour plot
of the $\chi_\nu^2$ values of the shape model fits for various
orientations of the spin pole. Figure~\ref{fig:shapechi2} shows the
result for $C=0.8$, which gave lower overall $\chi_\nu^2$ values than
the other values of $C$ that we tried. However the general
$\chi_\nu^2$ patterns are similar irrespective of the value of $C$.

\begin{figure*}
\plotone{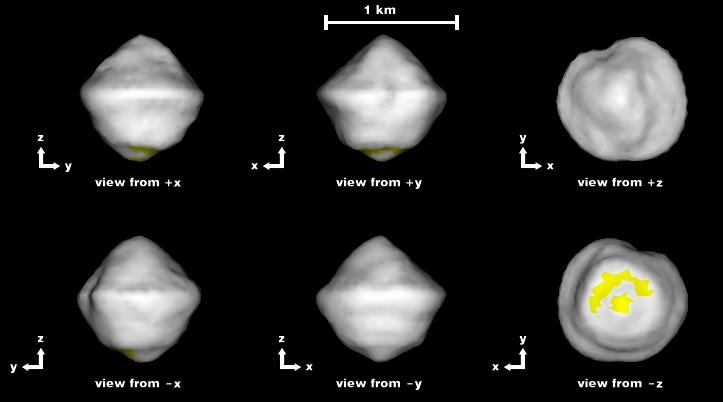}
\caption{Vertex shape model of the primary as seen along the three
  principal axes x, y, and z.  For principal axis rotation the spin
  axis is aligned with the z axis.  Yellow regions have radar
  incidence angles $> 60^\circ$ and hence are not well constrained.
  The shape model has 1000 vertices and 1996 triangular facets. The
  effective surface resolution is $\sim57$ m.}
\label{fig:primshape}
\end{figure*}

\begin{figure}
\plotone{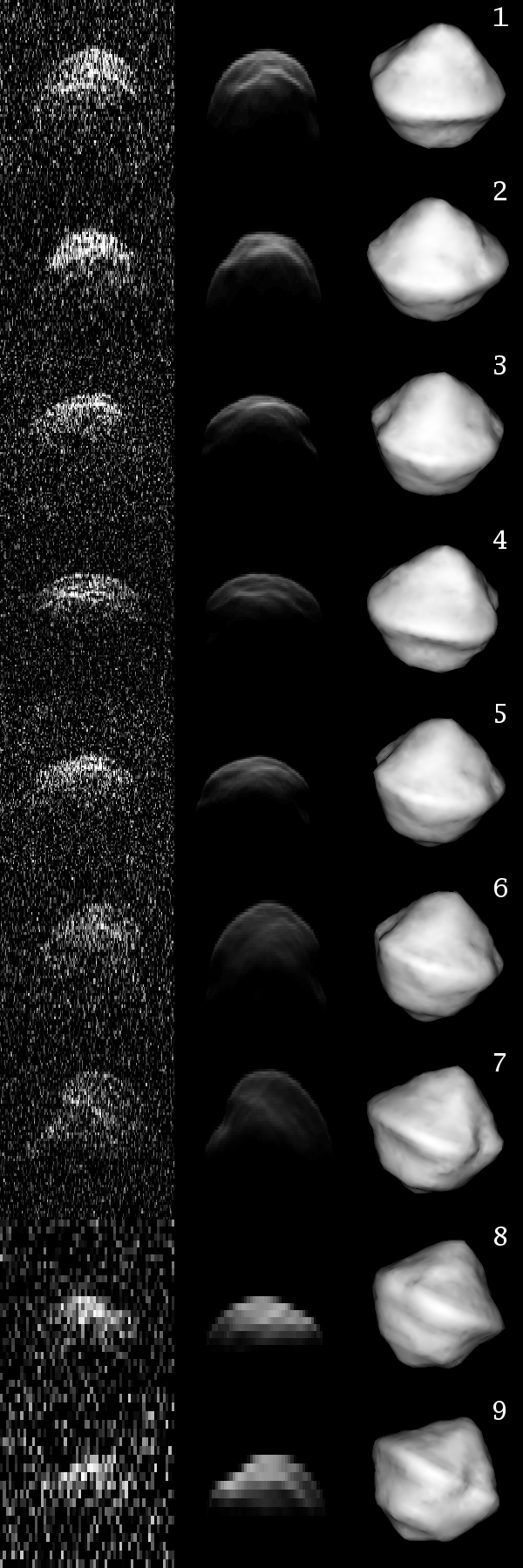}
\caption{Examples of images and fits for the primary. Each row (from
left to right) shows the observed image (single run), the
corresponding synthetic image generated using the shape model, and the
corresponding plane of sky view of the shape model.  The images were
obtained on (from top to bottom) September 10.43641, 10.47739,
11.41307, 11.46787, 13.37204, 15.33765, 15.38140, 18.30239, and
21.26607.  }
\label{fig:primfit}
\end{figure}

As explained in Section~\ref{sec:primshape}, we assumed the spin pole
to be aligned with the mutual orbit pole at $\lambda=294^\circ$ and
$\beta=78^\circ$.  The best-fit sidereal spin period is 2.7745 $\pm$
0.0007 h. Radar scattering $C=1.0$ yielded the shape model with the
lowest $\chi_\nu^2$. Figure~\ref{fig:primshape} shows the vertex shape
model produced under these assumptions for the spin pole and the value
of $C$, Table~\ref{tab:shapemodel} lists the associated parameters,
and Figure~\ref{fig:primfit} shows examples of the observed images and
the fits using this model.  The model shows a good general agreement
with the data.

An equatorial ridge similar to the one found on the 1999~KW4
primary~\citep{ostro06} is clearly seen. However the ridge is not so
regular and has a $\sim300$~m concavity on one side similar to
(341843) 2008 EV5~\citep{busch11}. An equatorial ridge is necessary to fit the
observed power profile behind the leading edge in the radar
images. The expected power profiles from models with and without
equatorial ridges are compared in \citet{busch11}.
The shape model shows another ridge-like structure forming
a ring around the south pole.

\begin{deluxetable}{lccc}
\tablewidth{0pt}
\tablecaption{Primary and secondary shape model parameters}
\tablehead{\colhead{Parameters} & \colhead{} & \colhead{Primary} & \colhead{Secondary}}
\startdata

Extents along                     &  x        & 0.992 $\pm$ 5\% & 0.379 $\pm$ 6\%\\
principal axes (km)               &  y        & 0.938 $\pm$ 5\% & 0.334 $\pm$ 6\%\\
                                  &  z        & 0.964 $\pm$ 5\% & 0.270 $\pm$ 6\%\\
\\
Surface area (km$^2$)             &           & 2.481 $\pm$ 10\%& 0.329 $\pm$ 12\%\\
\\
Volume (km$^3$)                   &           & 0.337 $\pm$ 15\%& 0.017 $\pm$ 18\%\\
\\
Moment of inertia ratios          & $A/C$ & 0.914 $\pm$ 10\%& 0.708 $\pm$ 10\%\\
                                  & $B/C$ & 0.946 $\pm$ 10\%& 0.888 $\pm$ 10\%\\
\\
Equivalent diameter (km)          &           & 0.863 $\pm$ 5\% & 0.316 $\pm$ 6\%\\
\\
DEEVE extents (km)                & x         & 0.899 $\pm$ 5\% & 0.377 $\pm$ 6\%\\
                                  & y         & 0.871 $\pm$ 5\% & 0.314 $\pm$ 6\%\\
                                  & z         & 0.821 $\pm$ 5\% & 0.268 $\pm$ 6\%\\
\\
Spin pole ($\lambda, \beta$) ($^\circ$) &       & (294, 78) & (294, 78) \\
\enddata

\tablecomments{The shape model of the primary consists of 1000
  vertices and 1996 triangular facets, corresponding to an effective
  surface resolution of $\sim57$ m.  The shape model of the secondary
  consists of 150 vertices and 296 facets; it has an effective surface
  resolution of $\sim$52 m.  Surface Area is
  the surface area of the shape model 
  measured at the model facet scale. 
  The moment of inertia ratios
  were calculated assuming homogeneous density.  $A$, $B$, and $C$ are
  the principal moments of inertia, such that $A<B<C$.  Equivalent
  diameter is the diameter of a sphere having the same volume as that
  of the shape model.  A dynamically equivalent equal volume ellipsoid
  (DEEVE) is an ellipsoid with uniform density having the same volume
  and moment of inertia ratios as the shape model.  The spin poles are
  assumed to be aligned with the mutual orbit pole.  }
\label{tab:shapemodel}
\end{deluxetable}

\subsection{Secondary Shape and Spin State}
\label{sec:rsecshape}

We found that including longitudinal libration in the secondary spin
model did not improve the shape model fits significantly, so we
adopted the shape model fit with no libration as the nominal shape
model. 
The non-detection of libration could either be because the libration
amplitude, which is predicted to be $\sim15$ m by \citet{naidu15}, is
less than the resolution of the images or the temporal and
longitudinal coverage of the secondary is insufficient.

\begin{figure}
\plotone{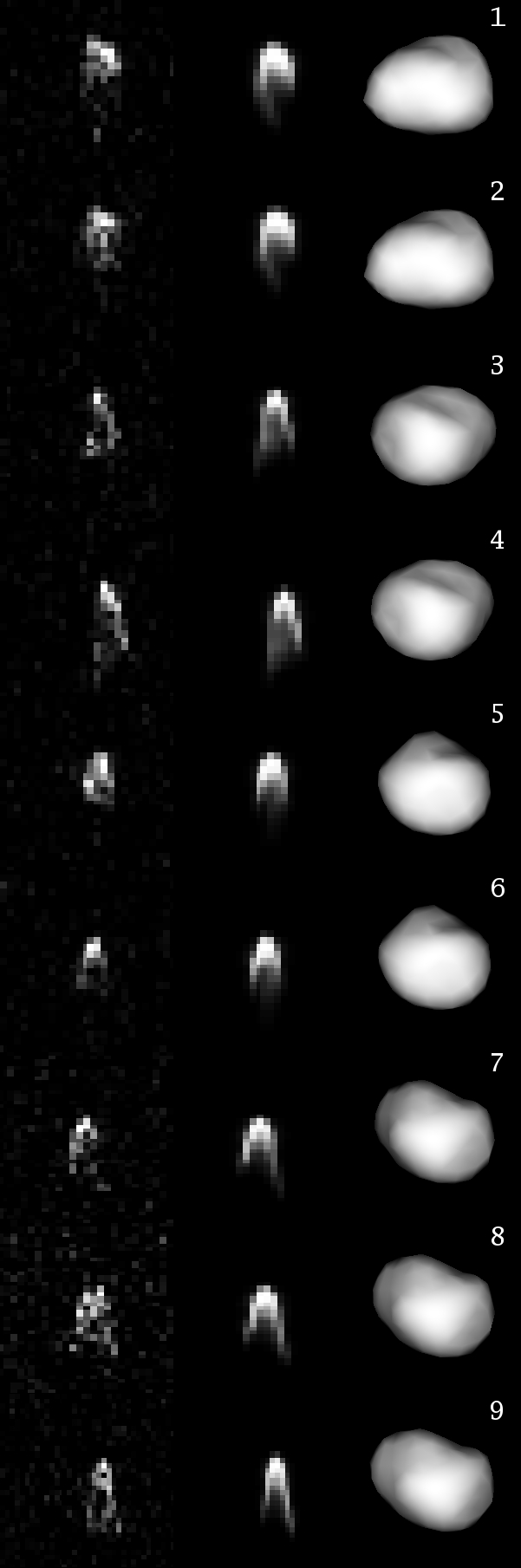}
\caption{Examples of images and fits for the secondary. Each row (from
left to right) shows the observed image (single run), the corresponding synthetic
image generated using the shape model, and the corresponding plane of
sky view of the shape model.  The images were obtained on (from
top to bottom) September 10.47192, 10.48012, 11.41307, 11.44529,
13.43063, 13.43896, 15.33765, 15.36390, and 15.39306.
}
\label{fig:secondaryfit}
\end{figure}

\begin{figure*}
\plotone{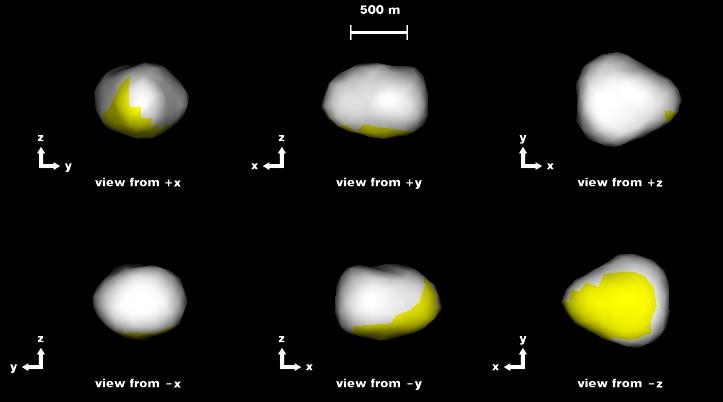}
\caption{Secondary shape model as seen along the three principal axes. Top
right view is along the positive spin axis. Yellow regions have radar
incidence angles $>60^\circ$ and hence are not well constrained.
The shape model has 150 vertices and 296 triangular facets. The effective
surface resolution is $\sim52$ m. }
\label{fig:secshape}
\end{figure*}

The best-fit sidereal spin period of the secondary is 1.77 $\pm$ 0.02
days. This is consistent with the radar derived mutual orbit period
suggesting that the secondary is spinning
synchronously. Figure~\ref{fig:secshape} shows the best-fit secondary
vertex shape model fit using this period,
Table~\ref{tab:shapemodel} lists the shape model parameters, and
Figure~\ref{fig:secondaryfit} shows some examples of the observed
images and the fits using this model.  There is good agreement between
the model and the data.
The secondary has a triangular pole-on-silhouette with dynamically
equivalent equal volume ellipsoid (DEEVE) dimensions of
$377\times314\times268$~m.

\subsection{Radar Scattering Properties}
\label{sec:rradarscat}
\begin{figure}
\plotone{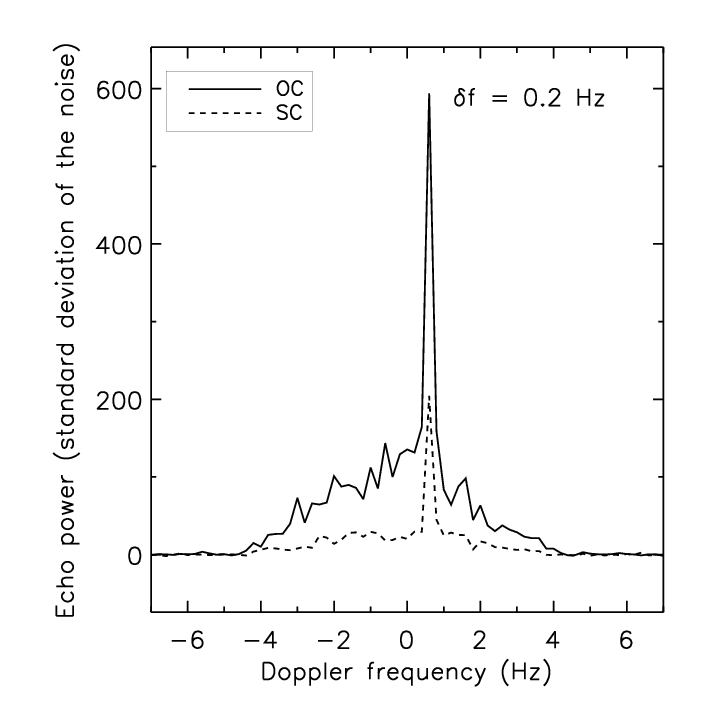}
\caption{Representative OC and SC CW spectra obtained at Arecibo on UT
  2008 September 11.483.  The broad component is due to the primary
  with 2.77 h spin period.  The narrow spike is due to the secondary
  with 1.77 d spin period.}
\label{fig:cw}
\end{figure}

Measurements of the OC radar albedo and circular polarization ratio
for the combined primary and secondary spectra
using the Arecibo data obtained in 2008 are listed in
Table~\ref{tab:rscattering}.  Their mean values are $0.179\pm0.02$ and
$0.265\pm0.03$, respectively, where the uncertainties are the standard
deviations of the individual measurements.  The mean value of circular
polarization ratio is close to the median value (0.26) for all NEAs
and is most consistent with the S- and C-class asteroids
\citep{benner08}. Figure~\ref{fig:cw} shows Arecibo OC and SC CW
spectra obtained on 2008
September 11.

Last two columns of Table~\ref{tab:rscattering} show the OC radar
albedos and circular polarization ratios for the power spectra
containing the estimated secondary contribution only.  Their mean
values are $0.174\pm0.05$ and $0.326\pm0.08$, respectively.  The radar
albedo of the secondary alone is equivalent to that of the
primary+secondary, suggesting that both components have identical
composition.  The polarization ratio of the secondary appears to be
more variable and greater than that of the primary, suggesting that
the secondary may be rougher than the primary at radar wavelength
scales.  However, the difference is within the 1 standard deviation of
the measurements, preventing a more definite conclusion.

\begin{deluxetable}{ccrrrr}
\tablewidth{0pt}
\tablecaption{Radar scattering properties}
\tablehead{\colhead{UT Date} & \colhead{Set} & \multicolumn{2}{c}{Prim.+Sec.} & \multicolumn{2}{c}{Secondary}\\
\colhead{yyyy-mm-dd} & \colhead{} & \colhead{$\Hat{\sigma}_{\rm OC}$} & \colhead{$\mu_C$} & \colhead{$\Hat{\sigma}_{\rm OC}$} & \colhead{$\mu_C$} } 
\startdata
2008-09-10 & 1 & 0.158   &   0.334  & 0.245    &  0.334 \\
2008-09-10 & 2 & 0.239   &   0.248  & 0.154    &  0.238 \\
2008-09-11 & 1 & 0.186   &   0.261  & 0.098    &  0.413 \\
2008-09-11 & 2 & 0.186   &   0.258  & 0.136    &  0.316 \\
2008-09-13 & 1 & 0.197   &   0.275  & 0.159    &  0.458 \\
2008-09-13 & 2 & 0.187   &   0.236  & 0.226    &  0.265 \\
2008-09-15 & 1 & 0.184   &   0.241  & 0.205    &  0.218 \\
2008-09-15 & 2 & 0.159   &   0.294  & 0.149    &  0.373 \\
2008-09-18 & 1 & 0.158   &   0.247  & 0.125    &  0.443 \\
2008-09-18 & 2 & 0.160   &   0.242  & 0.149    &  0.283 \\
2008-09-21 & 1 & 0.180   &   0.234  & 0.229    &  0.254 \\
2008-09-24 & 1 & 0.163   &   0.292  & 0.176    &  0.338 \\
2008-09-24 & 2 & 0.157   &   0.276  & 0.211    &  0.299 \\
\hline\\
Average    &   & 0.179   &   0.265  & 0.174    &  0.326\\ 
St. dev.   &   & 0.02    &   0.03   & 0.05     &  0.08 
\enddata

\tablecomments{Radar albedos $\Hat{\sigma}_{\rm OC}$ and circular
  polarization ratios $\mu_C$ of the primary and secondary combined
  (columns 3 and 4) and of the secondary alone (columns 5 and 6)
  measured on the basis of Arecibo data
  (Table~\ref{tab:observingsummary}).  Except for September 21, two
  measurements were available per day (distinguished by the index in
  the second column).}

\label{tab:rscattering}
\end{deluxetable}

\subsection{Mass Ratio, Component Masses, and Densities}
\label{sec:rmassratio}

\begin{figure}
\plotone{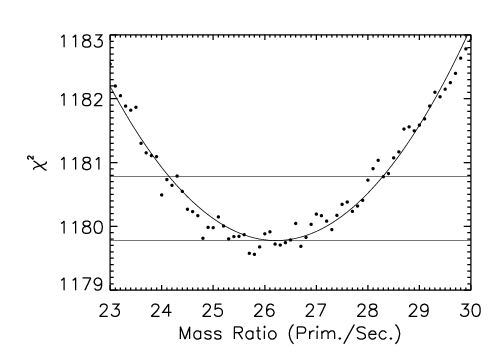}
\caption{Points show $\chi^2$ values of heliocentric orbit fits to
  optical and radar astrometric observations.  Radar astrometry
  includes two-way ranges to the system COM under various mass ratio
  assumptions. Solid curve shows the best fit parabola to the
  $\chi^2$'s. The horizontal lines show the minimum $\chi^2$ on the
  parabola and the $\chi^2$ corresponding to the $1\sigma$
  uncertainty, respectively. The minimum $\chi^2$ corresponds to a
  primary-to-secondary mass ratio of 26.2 $\pm$ 2.}
\label{fig:massratio}
\end{figure}

Direct estimation of the mass ratio using the method described in
Section~\ref{sec:massratio} yielded a mass ratio ($M_p/M_s$) of 26.2
$\pm$ 2.  This mass ratio corresponds to a reflex motion of the
primary of 98 $\pm$ 8 m, consistent with the estimate of 140 $\pm$ 40
m of \citet{margot02}, and with the apparent motion observed directly
in the images.  Figure~\ref{fig:massratio} shows a plot of the
$\chi^2$ values of the heliocentric orbit fits to the optical and
radar astrometry.  The latter uses two-way ranges to the system COM as
determined under various mass ratio assumptions as discussed in
Section~\ref{sec:massratio}.  Using this mass ratio we can apportion
the total mass of the system ($M_T$) to the two components. We find
the mass of the primary and the secondary to be
$4.656\pm0.43\times10^{11}$~kg and $0.178\pm 0.021\times10^{11}$~kg,
respectively.  Dividing the masses by the volumes of the corresponding
shape models, we find densities for the primary and secondary to be
$1381\pm244$~kg~m$^{-3}$ and $1047\pm230$ kg~m$^{-3}$, respectively,
where the largest source of uncertainty comes from the volume
determinations. The densities are similar, pointing towards a similar
composition and porosity.

\subsection{Primary Gravitational Environment}

\begin{figure*}
\plotone{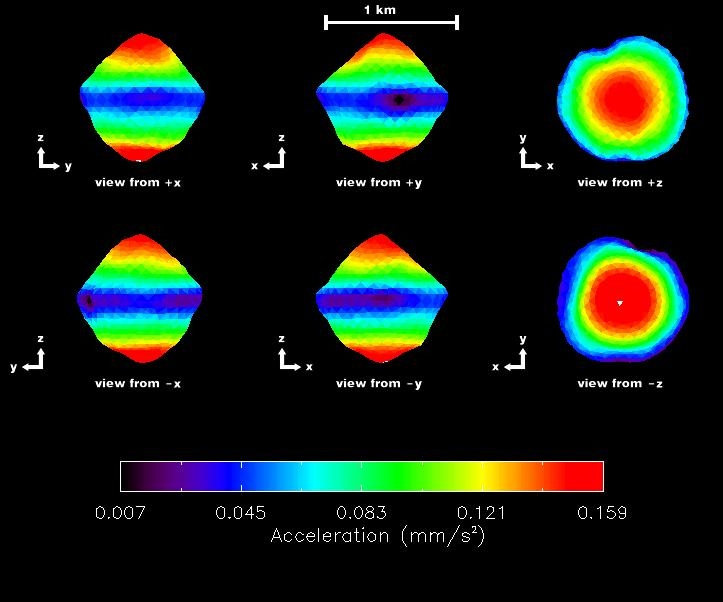}
\caption{This figure shows the magnitudes of the vector sum of
accelerations due to gravity and centrifugal accelerations computed at
the centers of the facets of the primary shape model. We assumed a
uniform density of 1381~$\rm{kg~m^{-3}}$, which was obtained in
Section~\ref{sec:rmassratio}, and a spin period of 2.7745~h. At the
equator, the values are close to zero, indicating that the magnitude
of centrifugal acceleration is almost equal to the magnitude of
acceleration due to the asteroid's mass. }
\label{fig:gravity}
\end{figure*}

\begin{figure*}
\plotone{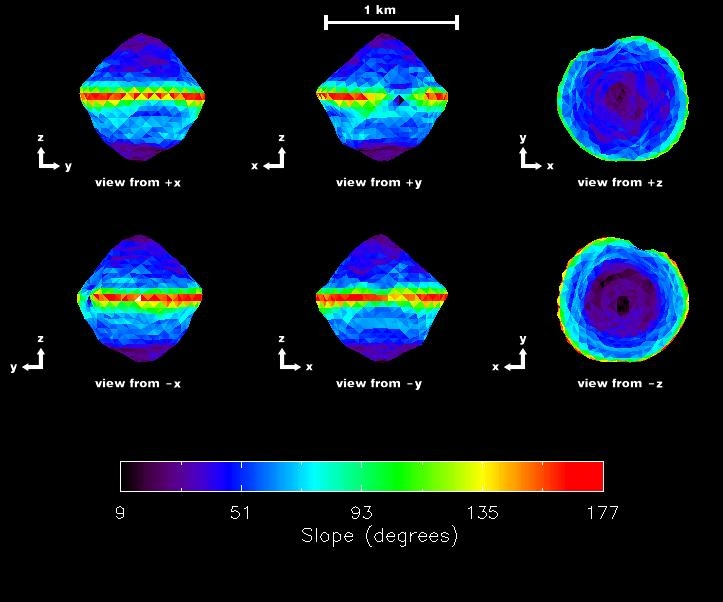}
\caption{This figure shows the gravitational slopes computed at the
centers of the facets of the primary shape model.
We assumed a uniform density of 1381~$\rm{kg~m^{-3}}$, obtained in
Section~\ref{sec:rmassratio}, and a spin period of
2.7745~h. Slopes vary from $\sim0^\circ$ at the poles and some
regions at the equator to close to $180^\circ$ at most regions at the
equator. Most regions at mid-latitudes have slopes between $40^\circ$
and $65^\circ$.\\ }
\label{fig:slopes}
\end{figure*}

The acceleration map on the surface of the primary shape model shows
that, for nominal values of mass, spin period, and shape parameters,
the net acceleration on the equatorial ridge is very close to zero
(Figure~\ref{fig:gravity}), which implies that the centrifugal
acceleration on the ridge almost cancels out the acceleration due to
the primary's mass. As we move to higher latitudes, and hence closer
to the spin axis, the magnitude of the centrifugal acceleration
decreases, causing the magnitude of the net acceleration to increase
and reach values up to 159~$\rm{\mu m~s^{-2}}$ at the poles.  This
value is about 1.6$\times$10$^{-5}$ times that on Earth. 

The gravitational slopes near the poles are close to zero
(Figure~\ref{fig:slopes}). Around the mid-latitudes, the slopes are
higher and most regions here have values between 40$^\circ$ and
65$^\circ$.
Regions on the equatorial ridge have slopes close to 180$^\circ$,
implying that the magnitude of centrifugal acceleration is greater
than the magnitude of acceleration due to mass. Inside the concavity
on the equatorial ridge the slopes are close to $0^\circ$. These
slopes provide clues to the mechanical properties of the asteroid
material. The implications are discussed in
Section~\ref{sec:discussion}.

\section{Discussion}
\label{sec:discussion}
\subsection{Primary shape and gravitational environment}
The primary shape is similar to shapes of some other
radar-characterized asteroids such as (66391) 1999~KW4, (136617)
1994~CC, (341843) 2008~EV5, (101955) Bennu,
etc \citep[respectively]{ostro06,brozovic11,busch11,nolan13}. This
commonly observed top-shaped structure is an indication that the
asteroid has undergone reshaping, most likely due to the spin-up of
the primary~\citep[e.g.,][]{harris09}. The shape and the gravitational
field provide clues about the mechanical properties of the material of
the primary. Figure~\ref{fig:slopes} shows that the gravitational
slopes around the mid-latitudes are mostly between 40$^\circ$ and
65$^\circ$. Some of these values are greater than the angle of repose
of sand on Earth which has values between 30$^\circ$ and 50$^\circ$.
A possible explanation of such high angles is that cohesive van der
Waals forces between the particles play an important role on the
surfaces of the asteroids as proposed by~\citet{scheeres10}.  These
cohesive forces could be comparable in magnitude to the ambient
gravitational force ~\citep{scheeres10}, resulting in much higher
effective angles of repose ($>50^\circ$) that the material can sustain
\citep[e.g.,][]{rognon08}. Figure~\ref{fig:slopes} shows that slopes
at the equator of the primary are $> 90^\circ$, implying that
centrifugal force is greater than the gravitational pull at the
equator. In the absence of other forces, this imbalance will cause
material to escape from the primary at the equator.  Cohesion between
particles could balance the excess centrifugal force and prevent such
an escape.  Nevertheless, the regions in the mid-latitudes having high
slopes might be devoid of fine grained material, as the material would
slide off to lower potential areas.  Some of the regions on the
equatorial ridge with slopes close to 180$^\circ$ might also be paths
through which material is shed off from the primary.  The slope values
are sensitive to the size, the density, and the spin period of the
asteroid. Scaling down the asteroid by $\sim5\%$ and keeping the mass
unchanged (effectively increasing its density by $\sim16\%$, which is
within the density uncertainty) yields slopes close to zero on most
regions at the equator and slopes lower than $45^\circ$ on most of the
surface of the asteroid. If tides and/or YORP spin down the asteroid,
there will be a global decrease in the slopes. A similar spin down
might have led to the overall low slopes seen on
2008~EV5~\citep{busch11}.

Assuming a grain density of 3000~kg~m$^{-3}$, which would be
appropriate for an S-type asteroid, the observed densities of the
primary and secondary can be explained by $\sim55\%$ and $\sim65\%$
porosity, respectively.  Dilation of cohesive materials during
avalanching flows seen in numerical simulations and laboratory
experiments \citep[e.g.,][]{alexander06, rognon08} could also explain
the high porosity needed to match the low densities of the primary and
the secondary.

The equatorial ridge has an approximately 300~m concavity on it. The
concavity could just be a void left over after the asteroid attained
its current shape or
it could be an impact crater. 
~\citet{jacobson11} hypothesized that a secondary fission event can
take place during the post-fission dynamics following the binary
formation process, and that one of the fragments may impact the
primary.  Secondary fission refers to the rotational fission of the
secondary as it is torqued by spin-orbit coupling while in a chaotic
rotation state~\citep{jacobson11,naidu15}.
Gravitational pull dominates the centrifugal force in the interior of
the concavity, so ponding of fine grained material transported from
higher latitudes can be expected inside the crater.

\subsection{Mutual Orbit}
\label{sec:dmutualorb}

The eccentricity of the mutual orbit, $e\approx0.019$, translates to a
variation of the primary-secondary distance of $2ae\approx 100$~m
during each orbit.  While this variation is detectable in the radar
data from 2008, which has a range resolution of $30$~m, it is barely
detectable in the radar data obtained in 2000, which has a range
resolution of 75~m.  Our determination of the longitude of pericenter
therefore relies on the 2008 data only.  Although we were not able to
fit an orbital precession rate, our method does not rule out
substantial pericenter precession during 2000-2008.  We performed
numerical simulations using the method developed by \citet{naidu15} to
estimate pericenter precession rates under various gravitational
perturbations: the non-spherical mass distribution of the primary
causes pericenter precession of about $90^\circ$/year, whereas the
non-spherical mass distribution of the secondary contributes about
$-15^\circ$/year. The combined effect causes the pericenter to precess
by about $75^\circ$/year in a prograde direction with respect to the
mutual orbit. Additionally, the gravitational perturbations from the
Sun cause the pericenter to precess by about $10^\circ$/year. The
combined effect of these three gravitational perturbations is a
secular apsidal precession rate of about $85^\circ$/year, but there
are significant short-term variations in the precession rate, making
detection of apsidal precession difficult. Gravitational perturbations
from planets and radiative forces from the Sun complicate the dynamics
further.

The mutual orbit normal (and the assumed primary and secondary spin
poles) is separated by about 5$^\circ$ from the heliocentric orbit
normal, which is common among binary NEAs and possibly indicative of
YORP obliquity evolution \citep{rubincam00}.

\subsection{Binary YORP}

Binary YORP is a radiative torque which is hypothesized to alter the
mutual orbit of synchronous binary systems~\citep{cuk05}.  A
synchronous satellite has a fixed leading and trailing side with
respect to the direction of its orbital motion, so an asymmetric
re-radiation from the surface of the satellite will lead to a net
torque on the mutual orbit. 
A potentially observable signature of such a torqued orbit is a
quadratic change in the mean anomaly of the
satellite~\citep{mcmahon10}. Detecting a quadratic change in mean
anomaly requires measurements of the mean anomaly on a minimum of 3
widely separated epochs. Additional measurements will be required to
model the complicated dynamics described in the previous
section. 2000~DP107 is a prime candidate for the detection of binary
YORP since it presents repeated opportunities for observations and has
already been observed in 2000 and 2008 by radar and in 2000, 2008,
2011, and 2013 by optical telescopes. \citet{mcmahon10} made a mean
anomaly drift rate prediction for 2000~DP107 by scaling the results
obtained from the radar-derived shape model of the satellite of
1999~KW4. Those predictions can now be updated using the secondary
shape model. Depending on the direction of the binary YORP torque, the
mutual orbit could either expand, contract, or remain unchanged. The
outcomes of these scenarios were studied in detail by
\citet{jacobson11}.  An expanding mutual orbit could lead to the
formation of asteroid pairs or an asynchronous satellite, whereas a
contracting mutual orbit could create a contact binary asteroid
\citep[e.g.,][]{taylor11}.
A contracting binary YORP torque could also be balanced by an equal
and opposite tidal torque implying a binary asteroid in a stable
equilibrium as hypothesized by \citet{jacobson11a}. Future
observations of this system may provide a detection of binary YORP
evolution.

\subsection{Formation and Evolution}
The normalized total angular momentum of a binary asteroid system
($J/J'$) provides clues to the formation mechanism of the system. In
this expression, $J$ is the total angular momentum and
$J'=\sqrt{GM_{\rm sys}R_{\rm eff}}$, where $M_{\rm sys}$ and $R_{\rm
  eff}$ are the total mass and equivalent radius of the binary system.
Ratios greater than 0.4 in NEAs are consistent with formation
of the binary by mass shedding due to spin-up of the parent
body~\citep{margot02,pravec07,taylor11}.  2000~DP107 has a separation
$a/R_{\rm p} \simeq 6.2 $ that is larger than most known binary NEAs
and a low eccentricity of 0.019, resulting in $J/J' \sim 0.49$.  This
is much larger than is necessary for spin fission.  In a tides-only
model, this large separation implies a rather weak primary, an old age
compared to the dynamical lifetime of NEAs, or the influence of
another mechanism such as binary YORP and/or YORP for increasing the
total angular momentum~\citep{taylor11}.

\section{Conclusion}
The radar observations of 2000 DP107 allowed us to produce shape
models of the primary and secondary, estimate their masses and
densities, compute the gravitational environment of the primary, and
estimate the mutual orbit parameters. The shape model and
gravitational environment of the primary provide important clues about
the material properties of the asteroid. The shape model of the
secondary can be used to estimate the evolution of the mutual orbit
under the binary YORP torque. Future radar and photometric
observations of the system may provide measurements of the evolution
of the mutual orbit.
The next radar and photometric observation opportunity is in 2016.

\section{Acknowledgements}

We thank Dan Scheeres and Seth Jacobson for useful discussions, and
the anonymous reviewer for excellent suggestions.  This material is
based upon work supported by the National Science Foundation under
Grant No. AST-1211581 and the National Aeronautics and Space
Administration under Grant No. NNX14AM95G.

\bibliographystyle{plainnat}


\begin{thebibliography}{33}
\providecommand{\natexlab}[1]{#1}
\providecommand{\url}[1]{\texttt{#1}}
\expandafter\ifx\csname urlstyle\endcsname\relax
  \providecommand{\doi}[1]{doi: #1}\else
  \providecommand{\doi}{doi: \begingroup \urlstyle{rm}\Url}\fi

\bibitem[Alexander et~al.(2006)Alexander, Chaudhuri, Faqih, Muzzio, Davies, and
  Tomassone]{alexander06}
Albert~W Alexander, Bodhisattwa Chaudhuri, AbdulMobeen Faqih, Fernando~J
  Muzzio, Clive Davies, and M~Silvina Tomassone.
\newblock Avalanching flow of cohesive powders.
\newblock \emph{Powder Technology}, 164\penalty0 (1):\penalty0 13--21, 2006.

\bibitem[{Benner} et~al.(2008){Benner}, {Nolan}, {Margot}, {Brozovic}, {Ostro},
  {Shepard}, {Magri}, {Giorgini}, and {Busch}]{benner08}
L.~A.~M. {Benner}, M.~C. {Nolan}, J.~{Margot}, M.~{Brozovic}, S.~J. {Ostro},
  M.~K. {Shepard}, C.~{Magri}, J.~D. {Giorgini}, and M.~W. {Busch}.
\newblock {Arecibo and Goldstone Radar Imaging of Contact Binary Near-Earth
  Asteroids}.
\newblock In \emph{AAS/Division for Planetary Sciences Meeting Abstracts \#40},
  volume~40 of \emph{Bulletin of the American Astronomical Society}, page 432,
  September 2008.

\bibitem[{Brozovi{\'c}} et~al.(2011){Brozovi{\'c}}, {Benner}, {Taylor},
  {Nolan}, {Howell}, {Magri}, {Scheeres}, {Giorgini}, {Pollock}, {Pravec},
  {Gal{\'a}d}, {Fang}, {Margot}, {Busch}, {Shepard}, {Reichart}, {Ivarsen},
  {Haislip}, {LaCluyze}, {Jao}, {Slade}, {Lawrence}, and {Hicks}]{brozovic11}
M.~{Brozovi{\'c}}, L.~A.~M. {Benner}, P.~A. {Taylor}, M.~C. {Nolan}, E.~S.
  {Howell}, C.~{Magri}, D.~J. {Scheeres}, J.~D. {Giorgini}, J.~T. {Pollock},
  P.~{Pravec}, A.~{Gal{\'a}d}, J.~{Fang}, J.-L. {Margot}, M.~W. {Busch}, M.~K.
  {Shepard}, D.~E. {Reichart}, K.~M. {Ivarsen}, J.~B. {Haislip}, A.~P.
  {LaCluyze}, J.~{Jao}, M.~A. {Slade}, K.~J. {Lawrence}, and M.~D. {Hicks}.
\newblock {Radar and optical observations and physical modeling of triple
  near-Earth Asteroid (136617) 1994 CC}.
\newblock \emph{Icarus}, 216:\penalty0 241--256, November 2011.
\newblock \doi{10.1016/j.icarus.2011.09.002}.

\bibitem[{Busch} et~al.(2011){Busch}, {Ostro}, {Benner}, {Brozovic},
  {Giorgini}, {Jao}, {Scheeres}, {Magri}, {Nolan}, {Howell}, {Taylor},
  {Margot}, and {Brisken}]{busch11}
M.~W. {Busch}, S.~J. {Ostro}, L.~A.~M. {Benner}, M.~{Brozovic}, J.~D.
  {Giorgini}, J.~S. {Jao}, D.~J. {Scheeres}, C.~{Magri}, M.~C. {Nolan}, E.~S.
  {Howell}, P.~A. {Taylor}, J.-L. {Margot}, and W.~{Brisken}.
\newblock {Radar observations and the shape of near-Earth ASTEROID 2008 EV5}.
\newblock \emph{Icarus}, 212:\penalty0 649--660, April 2011.
\newblock \doi{10.1016/j.icarus.2011.01.013}.

\bibitem[{{\'C}uk} and {Burns}(2005)]{cuk05}
M.~{{\'C}uk} and J.~A. {Burns}.
\newblock {Effects of thermal radiation on the dynamics of binary NEAs}.
\newblock \emph{Icarus}, 176:\penalty0 418--431, August 2005.
\newblock \doi{10.1016/j.icarus.2005.02.001}.

\bibitem[Funase et~al.(2014)Funase, Koizumi, Nakasuka, Kawakatsu, Fukushima,
  Tomiki, Kobayashi, Nakatsuka, Mita, Kobayashi, et~al.]{funase14}
Ryu Funase, Hiroyuki Koizumi, Shinichi Nakasuka, Yasuhiro Kawakatsu, Yosuke
  Fukushima, Atsushi Tomiki, Yuta Kobayashi, Junichi Nakatsuka, Makoto Mita,
  Daisuke Kobayashi, et~al.
\newblock 50kg-class deep space exploration technology demonstration
  micro-spacecraft procyon.
\newblock 2014.

\bibitem[{Gladman} et~al.(1996){Gladman}, {Quinn}, {Nicholson}, and
  {Rand}]{gladman96}
B.~{Gladman}, D.~D. {Quinn}, P.~{Nicholson}, and R.~{Rand}.
\newblock {Synchronous Locking of Tidally Evolving Satellites}.
\newblock \emph{Icarus}, 122:\penalty0 166--192, July 1996.
\newblock \doi{10.1006/icar.1996.0117}.

\bibitem[{Greenberg}(1981)]{gree81}
R.~{Greenberg}.
\newblock {Apsidal precession of orbits about an oblate planet}.
\newblock \emph{Astronomical Journal}, 86:\penalty0 912--914, June 1981.

\bibitem[{Harris} et~al.(2009){Harris}, {Fahnestock}, and {Pravec}]{harris09}
A.~W. {Harris}, E.~G. {Fahnestock}, and P.~{Pravec}.
\newblock {On the shapes and spins of rubble pile asteroids}.
\newblock \emph{Icarus}, 199:\penalty0 310--318, February 2009.
\newblock \doi{10.1016/j.icarus.2008.09.012}.

\bibitem[{Hudson}(1993)]{hudson93}
S.~{Hudson}.
\newblock {Three-dimensional reconstruction of asteroids from radar
  observations}.
\newblock \emph{Remote Sensing Reviews}, 8:\penalty0 195--203, 1993.

\bibitem[{Jacobson} and {Scheeres}(2011{\natexlab{a}})]{jacobson11}
S.~A. {Jacobson} and D.~J. {Scheeres}.
\newblock {Dynamics of rotationally fissioned asteroids: Source of observed
  small asteroid systems}.
\newblock \emph{Icarus}, 214:\penalty0 161--178, July 2011{\natexlab{a}}.
\newblock \doi{10.1016/j.icarus.2011.04.009}.

\bibitem[{Jacobson} and {Scheeres}(2011{\natexlab{b}})]{jacobson11a}
S.~A. {Jacobson} and D.~J. {Scheeres}.
\newblock {Long-term Stable Equilibria for Synchronous Binary Asteroids}.
\newblock \emph{\apjl}, 736:\penalty0 L19, July 2011{\natexlab{b}}.
\newblock \doi{10.1088/2041-8205/736/1/L19}.

\bibitem[{Magri} et~al.(2007){Magri}, {Ostro}, {Scheeres}, {Nolan}, {Giorgini},
  {Benner}, and {Margot}]{magri07}
C.~{Magri}, S.~J. {Ostro}, D.~J. {Scheeres}, M.~C. {Nolan}, J.~D. {Giorgini},
  L.~A.~M. {Benner}, and J.~L. {Margot}.
\newblock {Radar observations and a physical model of Asteroid 1580 Betulia}.
\newblock \emph{Icarus}, 186:\penalty0 152--177, January 2007.
\newblock \doi{10.1016/j.icarus.2006.08.004}.

\bibitem[{Margot} et~al.(2002){Margot}, {Nolan}, {Benner}, {Ostro}, {Jurgens},
  {Giorgini}, {Slade}, and {Campbell}]{margot02}
J.~L. {Margot}, M.~C. {Nolan}, L.~A.~M. {Benner}, S.~J. {Ostro}, R.~F.
  {Jurgens}, J.~D. {Giorgini}, M.~A. {Slade}, and D.~B. {Campbell}.
\newblock {Binary Asteroids in the Near-Earth Object Population}.
\newblock \emph{Science}, 296:\penalty0 1445--1448, May 2002.
\newblock \doi{10.1126/science.1072094}.

\bibitem[{Margot} et~al.(2015){Margot}, {Pravec}, {Taylor}, {Carry}, and
  {Jacobson}]{marg15AIV}
J.-L. {Margot}, P.~{Pravec}, P.~{Taylor}, B.~{Carry}, and S.~{Jacobson}.
\newblock {Asteroid Systems: Binaries, Triples, and Pairs}.
\newblock \emph{ArXiv e-prints}, 2015.

\bibitem[{McMahon} and {Scheeres}(2010)]{mcmahon10}
J.~{McMahon} and D.~{Scheeres}.
\newblock {Detailed prediction for the BYORP effect on binary near-Earth
  Asteroid (66391) 1999 KW4 and implications for the binary population}.
\newblock \emph{Icarus}, 209:\penalty0 494--509, October 2010.
\newblock \doi{10.1016/j.icarus.2010.05.016}.

\bibitem[{Mitchell} et~al.(1996){Mitchell}, {Ostro}, {Hudson}, {Rosema},
  {Campbell}, {Velez}, {Chandler}, {Shapiro}, {Giorgini}, and
  {Yeomans}]{mitchell96}
D.~L. {Mitchell}, S.~J. {Ostro}, R.~S. {Hudson}, K.~D. {Rosema}, D.~B.
  {Campbell}, R.~{Velez}, J.~F. {Chandler}, I.~I. {Shapiro}, J.~D. {Giorgini},
  and D.~K. {Yeomans}.
\newblock {Radar Observations of Asteroids 1 Ceres, 2 Pallas, and 4 Vesta}.
\newblock \emph{Icarus}, 124:\penalty0 113--133, November 1996.
\newblock \doi{10.1006/icar.1996.0193}.

\bibitem[Murray and Dermott(1999)]{murray99}
C.D. Murray and S.F. Dermott.
\newblock \emph{Solar System Dynamics}.
\newblock Cambridge University Press, 1999.
\newblock ISBN 9780521572958.
\newblock URL \url{http://books.google.co.uk/books?id=NY9iQgAACAAJ}.

\bibitem[{Naidu} and {Margot}(2015)]{naidu15}
S.~P. {Naidu} and J.-L. {Margot}.
\newblock {Near-Earth Asteroid Satellite Spins Under Spin-orbit Coupling}.
\newblock \emph{\aj}, 149:\penalty0 80, February 2015.
\newblock \doi{10.1088/0004-6256/149/2/80}.

\bibitem[{Naidu} et~al.(2013){Naidu}, {Margot}, {Busch}, {Taylor}, {Nolan},
  {Brozovic}, {Benner}, {Giorgini}, and {Magri}]{naid13}
S.~P. {Naidu}, J.-L. {Margot}, M.~W. {Busch}, P.~A. {Taylor}, M.~C. {Nolan},
  M.~{Brozovic}, L.~A.~M. {Benner}, J.~D. {Giorgini}, and C.~{Magri}.
\newblock {Radar imaging and physical characterization of near-Earth Asteroid
  (162421) 2000 ET70}.
\newblock \emph{Icarus}, 226:\penalty0 323--335, September 2013.
\newblock \doi{10.1016/j.icarus.2013.05.025}.

\bibitem[{Nolan} et~al.(2013){Nolan}, {Magri}, {Howell}, {Benner}, {Giorgini},
  {Hergenrother}, {Hudson}, {Lauretta}, {Margot}, {Ostro}, and
  {Scheeres}]{nolan13}
M.~C. {Nolan}, C.~{Magri}, E.~S. {Howell}, L.~A.~M. {Benner}, J.~D. {Giorgini},
  C.~W. {Hergenrother}, R.~S. {Hudson}, D.~S. {Lauretta}, J.-L. {Margot}, S.~J.
  {Ostro}, and D.~J. {Scheeres}.
\newblock {Shape model and surface properties of the OSIRIS-REx target Asteroid
  (101955) Bennu from radar and lightcurve observations}.
\newblock \emph{Icarus}, 226:\penalty0 629--640, September 2013.
\newblock \doi{10.1016/j.icarus.2013.05.028}.

\bibitem[{Ostro}(1993)]{ostro93}
S.~J. {Ostro}.
\newblock {Planetary radar astronomy}.
\newblock \emph{Reviews of Modern Physics}, 65:\penalty0 1235--1279, October
  1993.
\newblock \doi{10.1103/RevModPhys.65.1235}.

\bibitem[{Ostro} et~al.(2006){Ostro}, {Margot}, {Benner}, {Giorgini},
  {Scheeres}, {Fahnestock}, {Broschart}, {Bellerose}, {Nolan}, {Magri},
  {Pravec}, {Scheirich}, {Rose}, {Jurgens}, {De Jong}, and {Suzuki}]{ostro06}
S.~J. {Ostro}, J.~L. {Margot}, L.~A.~M. {Benner}, J.~D. {Giorgini}, D.~J.
  {Scheeres}, E.~G. {Fahnestock}, S.~B. {Broschart}, J.~{Bellerose}, M.~C.
  {Nolan}, C.~{Magri}, P.~{Pravec}, P.~{Scheirich}, R.~{Rose}, R.~F. {Jurgens},
  E.~M. {De Jong}, and S.~{Suzuki}.
\newblock {Radar Imaging of Binary Near-Earth Asteroid (66391) 1999 KW4}.
\newblock \emph{Science}, 314:\penalty0 1276--1280, November 2006.
\newblock \doi{10.1126/science.1133622}.

\bibitem[{Peale}(1969)]{peale69}
S.~J. {Peale}.
\newblock {Generalized Cassini's Laws}.
\newblock \emph{\aj}, 74:\penalty0 483, April 1969.
\newblock \doi{10.1086/110825}.

\bibitem[{Pravec} and {Harris}(2007)]{pravec07}
P.~{Pravec} and A.~W. {Harris}.
\newblock {Binary asteroid population. 1. Angular momentum content}.
\newblock \emph{Icarus}, 190:\penalty0 250--259, September 2007.
\newblock \doi{10.1016/j.icarus.2007.02.023}.

\bibitem[{Pravec} et~al.(1999){Pravec}, {Wolf}, and {{\v
  S}arounov{\'a}}]{pravec99}
P.~{Pravec}, M.~{Wolf}, and L.~{{\v S}arounov{\'a}}.
\newblock {How many binaries are there among the near-Earth asteroids?}
\newblock In J.~{Svoren}, E.~M. {Pittich}, and H.~{Rickman}, editors, \emph{IAU
  Colloq. 173: Evolution and Source Regions of Asteroids and Comets}, page 159,
  1999.

\bibitem[{Pravec} et~al.(2006){Pravec}, {Scheirich}, {Ku{\v s}nir{\'a}k}, {{\v
  S}arounov{\'a}}, {Mottola}, {Hahn}, {Brown}, {Esquerdo}, {Kaiser},
  {Krzeminski}, {Pray}, {Warner}, {Harris}, {Nolan}, {Howell}, {Benner},
  {Margot}, {Gal{\'a}d}, {Holliday}, {Hicks}, {Krugly}, {Tholen}, {Whiteley},
  {Marchis}, {Degraff}, {Grauer}, {Larson}, {Velichko}, {Cooney}, {Stephens},
  {Zhu}, {Kirsch}, {Dyvig}, {Snyder}, {Reddy}, {Moore}, {Gajdo{\v s}},
  {Vil{\'a}gi}, {Masi}, {Higgins}, {Funkhouser}, {Knight}, {Slivan}, {Behrend},
  {Grenon}, {Burki}, {Roy}, {Demeautis}, {Matter}, {Waelchli}, {Revaz},
  {Klotz}, {Rieugn{\'e}}, {Thierry}, {Cotrez}, {Brunetto}, and
  {Kober}]{pravec06}
P.~{Pravec}, P.~{Scheirich}, P.~{Ku{\v s}nir{\'a}k}, L.~{{\v S}arounov{\'a}},
  S.~{Mottola}, G.~{Hahn}, P.~{Brown}, G.~{Esquerdo}, N.~{Kaiser},
  Z.~{Krzeminski}, D.~P. {Pray}, B.~D. {Warner}, A.~W. {Harris}, M.~C. {Nolan},
  E.~S. {Howell}, L.~A.~M. {Benner}, J.~L. {Margot}, A.~{Gal{\'a}d},
  W.~{Holliday}, M.~D. {Hicks}, Y.~N. {Krugly}, D.~{Tholen}, R.~{Whiteley},
  F.~{Marchis}, D.~R. {Degraff}, A.~{Grauer}, S.~{Larson}, F.~P. {Velichko},
  W.~R. {Cooney}, R.~{Stephens}, J.~{Zhu}, K.~{Kirsch}, R.~{Dyvig},
  L.~{Snyder}, V.~{Reddy}, S.~{Moore}, {\v S}.~{Gajdo{\v s}}, J.~{Vil{\'a}gi},
  G.~{Masi}, D.~{Higgins}, G.~{Funkhouser}, B.~{Knight}, S.~{Slivan},
  R.~{Behrend}, M.~{Grenon}, G.~{Burki}, R.~{Roy}, C.~{Demeautis}, D.~{Matter},
  N.~{Waelchli}, Y.~{Revaz}, A.~{Klotz}, M.~{Rieugn{\'e}}, P.~{Thierry},
  V.~{Cotrez}, L.~{Brunetto}, and G.~{Kober}.
\newblock {Photometric survey of binary near-Earth asteroids}.
\newblock \emph{Icarus}, 181:\penalty0 63--93, March 2006.
\newblock \doi{10.1016/j.icarus.2005.10.014}.

\bibitem[Proakis and Salehi(2007)]{proakis2007}
J.G. Proakis and M.~Salehi.
\newblock \emph{Digital Communications}.
\newblock McGraw-Hill, 2007.
\newblock ISBN 9780072957167.
\newblock URL \url{http://books.google.com/books?id=HroiQAAACAAJ}.

\bibitem[{Rognon} et~al.(2008){Rognon}, {Roux}, {Naa\"im}, and
  {Chevoir}]{rognon08}
P.~G. {Rognon}, J.-N. {Roux}, M.~{Naa\"im}, and F.~{Chevoir}.
\newblock {Dense flows of cohesive granular materials}.
\newblock \emph{Journal of Fluid Mechanics}, 596:\penalty0 21--47, 2008.
\newblock \doi{10.1017/S0022112007009329}.

\bibitem[{Rubincam}(2000)]{rubincam00}
D.~P. {Rubincam}.
\newblock {Radiative Spin-up and Spin-down of Small Asteroids}.
\newblock \emph{Icarus}, 148:\penalty0 2--11, November 2000.
\newblock \doi{10.1006/icar.2000.6485}.

\bibitem[{Scheeres} et~al.(2010){Scheeres}, {Hartzell}, {S{\'a}nchez}, and
  {Swift}]{scheeres10}
D.~J. {Scheeres}, C.~M. {Hartzell}, P.~{S{\'a}nchez}, and M.~{Swift}.
\newblock {Scaling forces to asteroid surfaces: The role of cohesion}.
\newblock \emph{Icarus}, 210:\penalty0 968--984, December 2010.
\newblock \doi{10.1016/j.icarus.2010.07.009}.

\bibitem[{Taylor} and {Margot}(2011)]{taylor11}
P.~A. {Taylor} and J.-L. {Margot}.
\newblock {Binary asteroid systems: Tidal end states and estimates of material
  properties}.
\newblock \emph{Icarus}, 212:\penalty0 661--676, April 2011.
\newblock \doi{10.1016/j.icarus.2011.01.030}.

\bibitem[{Werner} and {Scheeres}(1997)]{werner97}
R.~A. {Werner} and D.~J. {Scheeres}.
\newblock {Exterior Gravitation of a Polyhedron Derived and Compared with
  Harmonic and Mascon Gravitation Representations of Asteroid 4769 Castalia}.
\newblock \emph{Celestial Mechanics and Dynamical Astronomy}, 65:\penalty0
  313--344, 1997.

\end{thebibliography}

\end{document}